\documentclass[12pt]{iopart}
\usepackage{xcolor}
\usepackage{graphicx}% Include figure files
\usepackage{dcolumn}% Align table columns on decimal point
\usepackage{bm}% bold math
\usepackage{multirow}
\usepackage[numbers,sort&compress,square]{natbib}
\usepackage{hyperref}% add hypertext capabilities
\usepackage{subcaption}
\usepackage{verbatim}
\usepackage[mathlines,running]{lineno}
\newcommand{\LSU}{Department of Physics, Louisiana State University, 202 Nicholson Hall
Baton Rouge, LA 70803 USA}
\newcommand{\CIERA}{Center for Interdisciplinary Exploration and Research in Astrophysics (CIERA), Department of Physics and Astronomy, Northwestern University, 1800 Sherman Ave, Evanston, IL 60201, USA}
\newcommand{\SUPA}{SUPA, School of Physics and Astronomy, University of Glasgow, Kelvin Building, University Ave, Glasgow G12 8QQ, UK}
\newcommand{\Syracuse}{School of Information Studies, Syracuse University, 343 Hinds Hall. Syracuse, NY 13210, USA}
\newcommand{\UWM}{Information School, University of Wisconsin--Madison, Helen C White Hall, 600 N Park Street, Madison, WI 53706, USA}
\newcommand{\CSF}{Nicholas and Lee Begovich Center for Gravitational-Wave Physics and Astronomy (GWPAC), Department of Physics, California State University Fullerton, Fullerton, 800 North State College Blvd, CA 92831, USA}
\newcommand{\EECS}{Electrical and Computer Engineering, Northwestern University, 2145 Sheridan Road, Evanston, IL 60208, USA}
\newcommand{\Adler}{Zooniverse, The Adler Planetarium, 1300 South Lake Shore Drive, Chicago, IL, 60605, USA}
\begin{document}

\title[Gravity Spy O3 glitches]{Discovering features in gravitational-wave data through detector characterization, citizen science and machine learning}
\author{
S~Soni$^{1}$, 
C~P~L~Berry$^{2,3}$,
S~B~Coughlin$^2$,
M~Harandi$^4$,
C~B~Jackson$^5$,
K~Crowston$^4$,
C~{\O}sterlund$^4$,
O~Patane$^6$,
A~K~Katsaggelos$^7$,
L~Trouille$^{8,2}$,
V-G~Baranowski$^9$, %Victor-Georges
W~F~Domainko$^9$, %Wilfried F.
K~Kaminski$^9$, %Kamil
M~A~Lobato~Rodriguez$^9$, %Miguel Angel
U~Marciniak$^9$, %Urszula
P~Nauta$^9$, %Peter 
G~Niklasch$^9$, %Gerhard
R~R~Rote$^9$, %Richard Ryan
B~Téglás$^9$, %Barbara
C~Unsworth$^9$ %Christine
and 
C~Zhang$^9$ %Chen 
}
\address{$^1$\LSU}
\address{$^2$\CIERA}
\address{$^3$\SUPA}
\address{$^4$\Syracuse}
\address{$^5$\UWM}
\address{$^6$\CSF}
\address{$^7$\EECS}
\address{$^8$\Adler}
\address{$^9$Gravity Spy}

\ead{siddharthsoni22@gmail.com}
\vspace{10pt}
\begin{indented}
\item[]
\end{indented}

\begin{abstract}
The observation of gravitational waves is hindered by the presence of transient noise (glitches). 
We study data from the third observing run of the Advanced LIGO detectors, and identify new glitch classes: \emph{Fast Scattering/Crown} and \emph{Low-frequency Blips}. 
Using training sets assembled by monitoring of the state of the detector, and by citizen-science volunteers, we update the Gravity Spy machine-learning algorithm for glitch classification. 
We find that Fast Scattering/Crown, linked to ground motion at the detector sites, is especially prevalent, and identify two subclasses linked to different types of ground motion. 
Reclassification of data based on the updated model finds that $\sim 27 \%$ of all transient noise at LIGO Livingston belongs to the Fast Scattering class, while $\sim 8\%$ belongs to the Low-frequency Blip class, making them the most frequent and fourth most frequent sources of transient noise at that site.
Our results demonstrate both how glitch classification can reveal potential improvements to gravitational-wave detectors, and how, given an appropriate framework, citizen-science volunteers may make discoveries in large data sets.
\end{abstract}


\section{Introduction}\label{sec:level1}

Advances in detector technology and in data collection drive discovery---access to new types of observations and more comprehensive data sets reveal previously unknown phenomena. 
A prime example is the operation of the Advanced LIGO~\cite{TheLIGOScientific:2014jea} and Advanced Virgo~\cite{TheVirgo:2014hva} detectors which has enabled the first direct observations of gravitational waves~\cite{Abbott:2016blz}. 
Gravitational waves provide a means to study merging neutron star and black hole binaries, which would be almost impossible otherwise, revolutionising our understanding of these astrophysical objects
\cite{TheLIGOScientific:2016htt,Abbott:2020gyp,LIGOScientific:2021qlt}.
Since the first gravitational-wave observation in 2015, the rate of detection has accelerated, reaching about one per week in the most recent observing run (O3)~\cite{Abbott:2020niy} which ran from April 2019 until March 2020. 
The wealth of data provided from ever-more sophisticated detectors enables scientific breakthroughs, but only if we can effectively use all of this data. 

Gravitational-wave detectors produce many forms of data---not only observations of gravitational waves from astrophysical sources, but also data on the operational state of the detector and the status of their environments~\cite{TheLIGOScientific:2016zmo,Nguyen:2021ybi,Fiori:2020arj}. 
How to intelligently use big data sets is a difficult challenge: because of their size, it is not always feasible for experts to carefully examine entire data sets.
One approach is crowd-sourcing analysis through citizen-science projects~\cite{Mendez:2008,Silvertown:2009,Bonney:2014}; here volunteers perform activities such as image classification at a scale impossible for a small science team. 
Alternatively, analysis may be automated through use of machine-learning algorithms~\cite{LHeureux:2017,Cuoco:2020ogp}; these can be trained to reproduce complicated operations enabling computers to quickly analyze data. 
Both citizen-science and machine-learning approaches have been successfully applied to a wide range of problems. 
However, in both citizen-science and machine-learning approaches it can be difficult to discover new features in the data---citizen-science volunteers, being non-experts, may not recognize interesting new phenomena without guidance, while machine-learning algorithms can struggle with unexpected features in the data. 
We examine the problem of identifying new transient noise features (\emph{glitches}) within data from the two LIGO gravitational-wave observatories in Livingston, Louisiana and Hanford, Washington.

\begin{figure}
    \centering
    \includegraphics[width=0.85\textwidth]{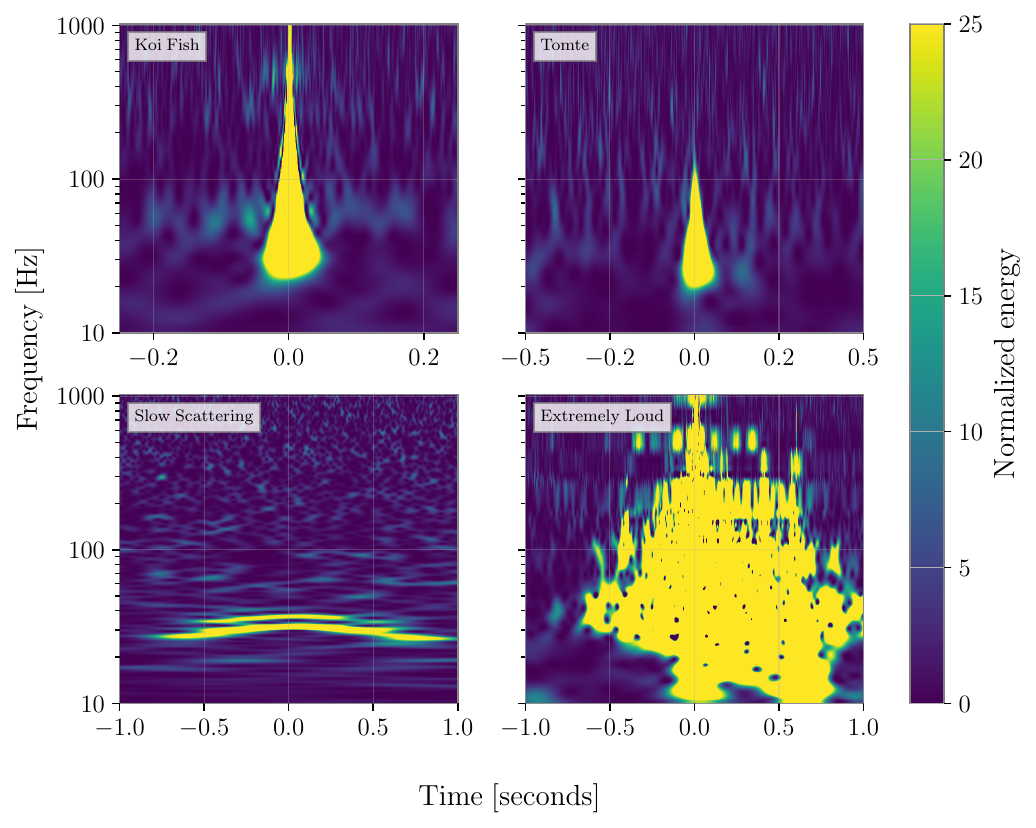}
    \caption{A selection of transient noise categories in gravitational-wave strain data. 
    These spectrograms illustrate the variety of time--frequency morphologies of glitches in LIGO data.
    \emph{Top left:} Koi Fish is short-duration broadband noise, usually with high SNR. 
    \emph{Top right:} Tomte is another short-duration noise category which occurred with a high rate at LIGO Livingston during O3.
    \emph{Bottom left:} Slow Scattering (Scattered Light) appears as long-duration arches between $10~\mathrm{Hz}$ and $120~\mathrm{Hz}$. 
    \emph{Bottom right:} Extremely Loud are very high SNR triggers that saturate the time--frequency plane and cause large drops in the astrophysical reach of the detector. 
    While the rate of Slow Scattering goes up with ground motion close to the detector, the other transient noise categories have not shown any environmental coupling.}
    \label{fig:glitches_combined}
\end{figure}

LIGO data quality is adversely affected by glitches. 
These short-duration bursts of noise can mimic or overlap a true gravitational-wave signal, complicating the process of detecting transient signals and inferring the properties of the source of the gravitational waves~\cite{Canton:2013joa,TheLIGOScientific:2017lwt,Pankow:2018qpo,Powell:2018csz,LIGOScientific:2019hgc,Chatziioannou:2021ezd}. 
Glitches can have environmental or instrumental origins~\cite{TheLIGOScientific:2016zmo,Nuttall:2018xhi,Cabero:2019orq,Davis:2021ecd}. 
There are a wide range of different glitch types; Figure~\ref{fig:glitches_combined} shows a time--frequency spectrogram, an Omega-scan~\cite{Chatterji:2004qg}, of common glitch types. 
As the detectors undergo commissioning, glitch classes can be eliminated as their sources are identified, but new classes may also arise (either because a change to a detector subsystem inadvertently introduces a glitch, or because improvements in detector sensitivity reveal a previously subdominant noise source).
For instance, before the start of O3, a number of upgrades were applied to the detectors, chief among them an increase in the laser power and the addition of squeezed light~\cite{Buikema:2020dlj}. 
These improvements increased the detectors' sensitivities by a factor of $2$ in the most sensitive frequency band compared to O2~\cite{Buikema:2020dlj,Abbott:2020niy}. 
However, this was accompanied by an increase in the glitch rate; at the Livingston Observatory the rate of glitches with signal-to-noise ratio (SNR) above $6.5$ quadrupled~\cite{Abbott:2020niy,Davis:2021ecd}. 
Identifying the origins of glitches is key to making the best use of gravitational-wave detector data.

Gravity Spy~\cite{Zevin:2016qwy} is a project that combines the power of citizen science and machine learning.\footnote{Gravity Spy Zooniverse project \href{http://gravityspy.org/}{gravityspy.org}} 
Using the Zooniverse platform, citizen-science volunteers are asked to sort glitches into a number of classes (including None of the Above and No Glitch). 
These classifications are used to build a training set used to train a convolutional neutral network (CNN)~\cite{Bahaadini:2017dqg}. 
The volunteers are trained through a series of levels, starting with classifying glitches the machine-learning algorithm calculates as confidently belonging to a subset of easily distinguishable classes, and progressing to classifying glitches which received a low probability of belonging to a given class by the algorithm.
The database of Gravity Spy classified glitches are routinely used in studies of the performance of the gravitational-wave detectors~\cite{Davis:2020nyf,Torres-Forne:2020eax,Cabero:2020eik,Jadhav:2020oyt,Soni:2020rbu,Krastev:2020skk}. 

The glitch classes included in Gravity Spy were originally defined by LIGO experts based on their studies of the detectors, with additional classes suggested by volunteers during early testing~\cite{Zevin:2016qwy,Bahaadini:2018git}. 
To aid in identifying new classes, Gravity Spy provides a search tool~\cite{Coughlin:2019ref} (available to both volunteers and experts) that identifies glitches in the data set that are similar to a nominal example.\footnote{Gravity Spy Tools \href{https://gravityspytools.ciera.northwestern.edu/}{gravityspytools.ciera.northwestern.edu/}} 
The similarity search uses the machine-learning algorithm's feature space to identify morphologically similar glitches~\cite{Coughlin:2019ref}, where similarity is measured using the cosine distance between glitches in a feature space that is constructed by the CNN~\cite{Bahaadini:2018nkh}. 
Here we study how Gravity Spy enabled the discovery of two new prevalent glitch classes---\emph{Fast Scattering}/\emph{Crown} and \emph{Low-frequency Blips}---during the O3 run of Advanced LIGO.

Assessing LIGO data quality is a time-consuming process. 
As early as October--November 2018, commissioners at LIGO Livingston began to notice a new type of daytime noise that appeared as short-duration arches in time--frequency spectrograms; this came to attention of detector-characterization experts in March 2019, and the new glitch would eventually become known as Fast Scattering/Crown~\cite{alog:44803,dcc:daytime_noise}.  
%As opposed to the long-duration Slow Scatter, Fast Scattering/Crown arches occur with shorter duration. 
%A comparison between Slow and Fast Scattering noise is shown in Figure~\ref{fig:slow_fast}. 
%This transient noise was new and was not observed in the earlier runs of LIGO. 
Similarly, during O3 detector-characterization experts noticed a new population of short-duration, broadband noise transients that were similar to the established Blip class of glitches, but with a lower maximum frequency. 
This category came to be known as Low-frequency Blips~\cite{dcc:gspy_sidd}. 
We have identified a potential correlation between Fast Scattering/Crown glitches and instrument vibration, while for Low-frequency Blips (as for Blips~\cite{Cabero:2019orq}), the cause still remains a mystery. 
Both types of transient noise were prevalent in the O3 data, impacting the data quality in different ways. 
Low-frequency Blips, due to their morphology, can be confused for short-duration gravitational-wave transients, such as signals from high-mass compact binaries hindering detection~\cite{Davis:2021ecd,LIGOScientific:2021hoh}. 
%These transients thus, can trigger the detection pipeline resulting in false positives. 
Fast Scattering/Crown glitches on the other hand, negatively impacted the overall sensitivity of the detectors, and also overlapped with multiple detections, necessitating their excision from the data before the sources of the signals could be inferred~\cite{Abbott:2020niy,LIGOScientific:2021qlt}. 
The output of our work is a classification of glitches from O3 that can be used in future studies of LIGO data.

Identification of new glitch classes is an important step in improving the LIGO detectors. 
We show how Gravity Spy has enabled the independent identification of a new glitch class via two complementary methods. 
The first is the traditional approach of carefully studying the state of the LIGO instruments and their environments to identify potential noise sources~\cite{Nguyen:2021ybi}.
The second is through Gravity Spy citizen-science volunteers exploring the data from the detectors to identify new patterns. 
Despite the citizen-science volunteers not having access to auxiliary information about the instruments or environment, an analogous new glitch type was found by both methods, highlighting the potential of citizen-science volunteers for making discoveries based upon their own in-depth investigations. 
We show how the updated glitch classification is applied to O3 data, demonstrate how this new class is related to ground motion at the observatory sites, and discuss how future discoveries of new transient noise features could be facilitated.

The success of Gravity Spy in enabling the identification of the Fast Scattering/Crown glitch class may serve as a case study for how citizen-science volunteers can make discoveries if supplied with appropriate training, analysis tools and expert feedback. 
Given the increasing prevalence of large scientific data sets, such citizen-science investigations could lead to breakthroughs in many other fields.\footnote{Since 2007, the Zooniverse platform has launched over $200$ projects across disciplines from astronomy to zoology, cancer research to climate science, and the arts and the humanities.}

\section{Methods}\label{sec:Methods}

In Sec.~\ref{sec:detchar} we introduce how glitches are identified in LIGO data, and highlight the need to assemble training sets for new glitch classes. 
In Sec.~\ref{sec:expert} we provide examples of how new glitches are identified by LIGO detector-characterization studies, while in Sec.~\ref{sec:volunteer} we discuss the equivalent practices of the citizen-science volunteers. 
During O3, both approaches lead to the convergent identification of a new glitch class linked to ground-motion at the detector sites, named Fast Scattering by the LIGO experts and Crown by the volunteers; we concentrate on this class, while also discussing a second class identified by the experts, Low-frequency Blips. 

\subsection{Detector characterization \& Gravity Spy}\label{sec:detchar}

Detector characterization is an umbrella term for monitoring data quality, investigating the origin of noise in the detector, enhancing the detector performance by mitigating noise, and validating gravitational-wave signals~\cite{Davis:2021ecd}. 
Apart from Gravity Spy, several other software tools such as {Hveto}~\cite{hveto}, {gwpy}~\cite{gwpy}, {gwdetchar}~\cite{gwdetchar}, {Omicron}~\cite{Robinet:2015om,Robinet:2020lbf} and iDQ~\cite{Essick:2020qpo} are used in detector characterization for studying transient noise properties. 
The output from noise-analysis tools as well as the state of various subsystems of the detector, is shown on a set of webpages known as LIGO Summary pages~\cite{gwsumm}. 
Detector characterization is essential to gravitational-wave astronomy.

Potential glitches in the LIGO data are found by event trigger generators that identify  excess power in the data stream. 
All the noise transients analyzed in this paper were detected by the Omicron algorithm~\cite{Robinet:2015om,Robinet:2020lbf}. 
%These transients processed by Omicron are also called triggers, and from here on, we will use glitch and trigger interchangeably. 
The Omicron algorithm annotates each of these detected triggers with specific characteristics such as event time, peak frequency, central frequency and SNR. 
Other than these characteristics, the time--frequency structure of a glitch is the most important feature that separates it from other transient noise sources. 
The glitch morphology of the trigger can be visualized in a time--frequency spectrogram commonly known as an Omega-scan~\cite{Chatterji:2004qg, gwdetchar}. 
These Omega-scans are used frequently in data-quality studies to establish potential noise correlations between different parts of the detector~\cite{alog:transmon}. 
Examples of the range of morphologies are shown in Figure~\ref{fig:glitches_combined};  these examples are some of the most common glitch classes found in O3. 
Different sources of transient noise saturate different areas of the time--frequency plane. 
These distinguishing patterns can be used to categorize glitches. 

\subsubsection{Machine-learning algorithm \& training}\label{sec:CNN}

Gravity Spy uses a CNN to classify glitches into categories~\cite{Zevin:2016qwy,Bahaadini:2017dqg}.  
Given an image, a trained CNN architecture will recognize different aspects in the image and assign varying degrees of weights and biases to these predictive features. 
For Gravity Spy, each input image is a combination of four time--frequency spectrograms of $0.5~\mathrm{s}$, $1.0~\mathrm{s}$, $2.0~\mathrm{s}$ and $4.0~\mathrm{s}$ duration centered on the same transient~\cite{Zevin:2016qwy,Bahaadini:2017dqg}.
Some of the noise categories such as the Blip are sub-second in duration whereas others like Slow Scattering last $3$--$4~\mathrm{s}$. 
This concatenation of multiple duration images ensures that relevant features for each class are captured by the training set. 
For each image fed to the model, the algorithm will calculate a probability score (confidence) for each class, and the image is assigned the class with the highest associated confidence.

The CNN architecture of Gravity Spy~\cite{Bahaadini:2017dqg} consists of multiple layers.
There are four convolutional layers to extract features, each followed by a max-pooling and a rectified linear unit (ReLU) activation layer. 
After these layers, there are a fully connected layer and a final softmax layer. 
The number of nodes in the final softmax layer is equal to the number of glitch classes. 
The output score of the softmax layer is 
\begin{equation}
    o_{c}^{i} = \frac{\exp({w_{c}}\cdot x)}{\sum_{c=1}^{C} \exp({w_{c}}\cdot {x})}, 
\end{equation}
where $o^{c}_{i}$ represents the probability score that the image $i$ belongs to the glitch category $c$, $C$ is the total number of glitch classes, and $w$ represents the weight vector that connects the output of the previous layer to the $c$-th node in the softmax layer.  
Weights are neural network parameters that represent the strength of connection between different nodes. 
This output score is used to assign a probability distribution across glitch classes~\cite{Zevin:2016qwy}. 
While training, the CNN model optimizes a cross entropy loss function
\begin{equation}
    \mathcal{L} = -\sum_{i=1}^{N}\sum_{c=1}^{C}y^{i}_{c}\log o^{i}_{c},
\end{equation}
where $y^{i}_{c}$ is an indicator function for the true label of the image $i$ that equals $1$ if the image belongs to class $c$ and equals $0$ for all other classes, and $N$ is the total number of training samples. 

%To optimize this loss function, Gravity Spy uses Adadelta optimizer \cite{2012arXiv1212.5701Z}. 

%% OLD TEXT FOROM ORIGINAL VERSION OF THE PAPER
%CNNs are widely used for the analysis of visual imagery and are recognized as the primary choice of algorithm for tasks related to computer vision. 
%Gravity Spy uses Keras Deep Learning Library with TensorFlow backend.
%Given an image, a trained CNN architecture will recognize different aspects in the image and assign varying degrees of weights and biases to these predictive features. 
%The model architecture of Gravity Spy contains two convolutional and two fully connected layers. 
%The input shape of first CNN layer match the  dimension of the image input and the dimension of the last Dense layer match the number of distinct classes for this classification. 
%Softmax is the choice of activation function for the fully connected output layer whereas ReLu is used by the first layers in the model. 
%For each image fed to the model, the algorithm will calculate a probability score (confidence) for each class. 
%This probability score is given by the output of softmax layer as
%\begin{equation}
%    o_{c}^{i} = \frac{\exp({w_{c}}\cdot x)}{\sum_{c=1}^{C} \exp({w_{c}}\cdot {x})},
%\end{equation}
%where $o_{c}^{i}$ is the probability of $i$-th image for class $c$, $C$ is the total number of classes, $w_{c}$ is the weight vector connected to the $c$-th node in the softmax layer and $x$ is the input given to the softmax layer.
%The image is then assigned a class with the highest associated probability.

Key to the performance of the CNN is the training set.
The Gravity Spy training set is composed of time--frequency spectrogram images of different glitches that have been selected by experts and volunteers to exemplify the various classes. 
The total number of training images for each category is a function of how frequently it appears in the data. 
Blip and Slow Scattering, for example, have more training data as they occur with a relatively higher rate: Blips occurred at a rate of about $2$ per hour in each of the two LIGO detectors during the second observing run (O2)~\cite{Davis:2021ecd}. 
%In addition, this \change{overabundance of common glitches} was amplified by the structure of the Gravity Spy workflows: the earliest levels contained fewer options in order to enhance volunteer training, but this means that those classes received more volunteer annotations than classes that were introduced later in the training. 
%For instance, Level 1 only has Blips and Whistles, which led to more of the training set being Blips and Whistles. 
The previous training set contains a total of $7932$ training images distributed among $22$ classes~\cite{Bahaadini:2018git}; the Gravity Spy model trained on this data set was used during the first two LIGO observing runs as well as during O3.\footnote{The old $22$ glitch classes are: 1080 Lines, 1400 Ripples, Air Compressor, Blip, Chirp (simulated gravitational-wave signals), Extremely Loud, Helix, Koi Fish, Light Modulation, Low-frequency Burst, Low-frequency Line, Paired Dove, Power Line, Repeating Blips, Scattered Light (which we will rename to Slow Scattering), Scratchy, Tomte, Violin Mode Harmonic, Wandering Line, Whistle, No Glitch (for when no glitch is visible), and None of the Above (for glitches that do not match the other categories).} 

Here, we discuss updating the Gravity Spy training set to account for changes to the data quality in O3. 
Our new training set includes two new glitch classes: Fast Scattering/Crown and Low-frequency Blips. 
Adding the new glitch classes enables us to use the machine-learning algorithm to identify examples of these glitches, so that they can be studied in detail.
Following O3, we retrained the CNN using the updated training sets, and reclassified the data set with this new model. Statistics for the most common glitch classes as identified by our old and new machine-learning models are given in Table~\ref{tab:top}.

\subsection{Overview of O3 data quality \& expert investigations}\label{sec:expert}

During O3, Gravity Spy classified $195445$ and $125256$ Omicron triggers from LIGO Livingston and LIGO Hanford, respectively, into one of the $22$ pre-O3 classes with confidence above $90\%$. 
In comparison to LIGO Hanford, the Livingston Observatory experiences a higher rate of glitches due to its better sensitivity in the $\sim10$--$60~\mathrm{Hz}$ band and worse ground-motion conditions near the site.
Noise due to light scattering dominated at both Livingston and Hanford with $40.3\%$ and $46.6\%$ of the triggers being classified as \emph{Scattered Light} at the two sites respectively (see Table~\ref{tab:top}). 
This is not surprising as ground motion is a persistent source of noise at both of the observatories. 
During high ground motion, the vibrating hardware in the detector can reflect any stray light incident on it back into the main beam; 
the relative motion between the vibrating surface and the main optic's mirrors introduces a time-varying phase to this stray light which shows up as noise~\cite{Accadia:2010zzb,Ottaway:2012oce,Valdes:2017xce,Soni:2020rbu}.
At Livingston, the glitch categories Tomte and None of the Above are the most dominant after Scattered Light while at Hanford, Low-frequency Burst and Extremely Loud are the second and third most frequent source of transient noise respectively.
The frequency of different glitch classes varies both between the two observatories and between observing runs.

Two major clues pointed towards the need to recognize at least one new glitch category during O3:
\begin{enumerate}
    \item A high number of None of the Above at Livingston (around $12\%$ of all the classified triggers; see Table~\ref{tab:top}).
    \item There were triggers classified as {Scattered Light} with a short duration ($0.25$--$0.3~\mathrm{s}$) glitch morphology compared to the prototypical scattering arch shown in the bottom left of Figure~\ref{fig:glitches_combined}. 
\end{enumerate}
The SNR and duration distribution of the Livingston glitches classified as Scattered Light hinted towards the presence of another population, one with a shorter duration and lower SNR than the conventional Scattered Light glitches~\cite{Soni:2020rbu}. 
The time--frequency spectrograms of a number of {None of the Above} glitches appeared similar to the new population of {Scattered Light}. 
The differences from the traditional Scattered Light suggest a separate physical origin for this new population.
A time comparison showed that days with a high number of {None of the Above} at Livingston were also the days with high ground motion, which is the main contributor of noise due to light scattering. 
These investigations pointed towards the presence of a new class of glitches with a coupling to ground motion similar to, but distinct from, the previously observed Scattered Light. 
This new glitch category came to be known within LIGO as \emph{Fast Scattering} due to its short-duration arches, while the longer duration Scattered Light became known as \emph{Slow Scattering}. 

\begin{figure}
    \centering
    \includegraphics[width=0.75\textwidth]{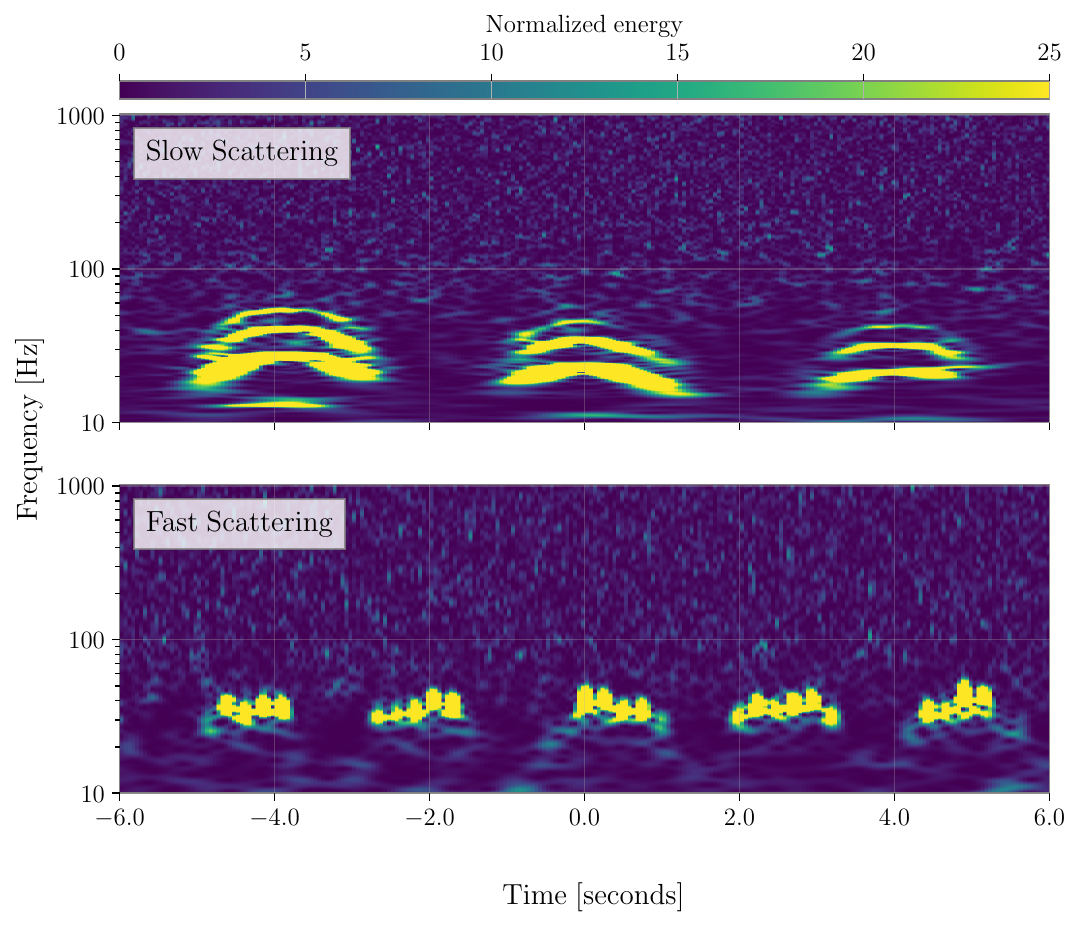}
    \caption{
    Comparison of glitches caused by light scattering. 
    \emph{Top:} Slow Scattering shows up as long duration arches and is correlated with an increase in ground motion in the earthquake ($0.03$--$0.1~\mathrm{Hz}$) and microseismic ($0.1$--$0.3~\mathrm{Hz}$) band. 
    \emph{Bottom:} Fast Scattering often consists of clusters of arches with a higher frequency, and is more common during an increase in anthropogenic band ground motion ($1$--$6~\mathrm{Hz}$). 
    The noise transients shown here are during the passing of a train close to the Livingston site when Fast Scattering triggers occur in clusters.}
    \label{fig:slow_fast}
\end{figure}

Fast Scattering glitches are subsecond-duration noise transients that affect LIGO data quality in the $20$--$60~\mathrm{Hz}$ range. A comparison of Slow Scattering and Fast Scattering is shown in Figure~\ref{fig:slow_fast}. 
The rate of Fast Scattering is found to be strongly correlated with ground motion in $0.1$--$0.3~\mathrm{Hz}$ and $1$--$6~\mathrm{Hz}$ frequency bands~\cite{dcc:noise-sprint}. 
Human activity, thunderstorms, logging and trains (in the case of Livingston) near the site are major causes of ground motion in this band. 
Fast Scattering is more common at Livingston than at Hanford due to differences in ground motion as well as the detector sensitivity~\cite{alog:ashley}. 

In addition to Fast Scattering, another addition to the Gravity Spy classification set is the Low-frequency Blip. 
Blips are sub-second duration broadband glitches that show up frequently at both the LIGO detectors~\cite{Cabero:2019orq}; they are yet to show any environmental coupling. 
Due to their morphology, these transients are considered especially harmful to search for high-mass binaries~\cite{LIGOScientific:2018mvr,Salemi:2019ovz,Davis:2020nyf}. 
Efforts to understand their origin have largely been unsuccessful though many sources have been ruled out~\cite{alog:blipH1,alog:blip_statistics,dcc:blip_LHO_Adrian}. 
During O3, we noticed many glitches with duration similar to Blips but with much lower peak frequencies were being classified as Blips or None of the Above. 
Blips usually have the frequency range $\sim32$--$2048~\mathrm{Hz}$, while the majority of transients from the new population are in the band $\sim32$--$128~\mathrm{Hz}$.
Therefore, we created a new Gravity Spy category for these glitches.
% We are yet to discover an instrumental or environmental origin for these glitches.
% Examples of both Blip and Low-frequency Blip glitch categories are shown in Figure~\ref{fig:blip_lowf_blip}. 

\subsubsection{Expert-selected training set}

To update the Gravity Spy model to include O3 glitches, training examples were needed. 
Our search for Fast Scattering began with finding days in O3 with high ground motion in the $1$--$6~\mathrm{Hz}$ frequency band at Livingston. 
We concentrated on Livingston as this noise is more prevalent there. 
Next, for these days we randomly selected $400$ triggers classified as Scattered Light by the old model with quality factor ($Q$-value) between $8$ and $14$.
The $Q$-value is defined as the ratio of the central frequency $f_{0}$ of the glitch to its bandwidth $\sigma_f$,
\begin{equation}
    Q = \frac{f_{0}}{\sigma_{f}},
\end{equation}
and is a useful metric in distinguishing glitch classes.
The $Q$-value intervals for Fast Scattering and Slow Scattering are disjoint due to the difference in glitch morphology. 
These $400$ triggers formed the Fast Scattering training data. 
For Low-frequency Blips, we selected $630$ glitches classified as Blips during O3 with peak frequency between $10~\mathrm{Hz}$ and $50~\mathrm{Hz}$. 
Low-frequency Blip and Tomte glitches affect the LIGO data quality in a similar frequency band of $\sim 30$--$120~\mathrm{Hz}$.
To ensure that the machine-learning classifier does not get confused between these two classes, we also increased the number of Tomte glitches in the training set. 
We found new examples of Tomte glitches by randomly sampling $300$ glitches from the O3 data, and added these to the training set. 
We also added $150$ examples of Slow Scattering from Hanford, as some examples of Slow Scattering from Hanford were being misclassified as Extremely Loud by the old model. 
All the data added to the training set were previously classified with confidence above $0.9$.
As a final change, we removed the training data associated with the None of the Above category. 
None of the Above was primarily intended to enable our citizen-science volunteers to flag new glitch classes; having now established the major classes for the O3 set it was not needed (although it would be reintroduced for future observing runs). 
The new O3 training data set contains $9631$ training examples distributed over $23$ classes. 

The Gravity Spy algorithm was trained on this data set for a total of $100$ epochs, with a $87.5$:$12.5$ split between training and validation set. 
At the end of the training, the training and validation set accuracies were $99.9$\% and $98.8$\% respectively.

\subsection{Volunteer methods}\label{sec:volunteer}

Gravity Spy citizen-science volunteers have the ability to suggest new glitch classes.
To better understand the volunteers' process for submitting a new glitch proposal, we conducted interviews with $9$ Gravity Spy participants. 
Six one-hour interviews were carried out via video conference.
%These were recorded for analysis. 
Three other interviews were conducted via email. 
The interviews provided insights about work practices e.g., using computational tools, and interactions e.g., collaboration with other Gravity Spy participants when investigating and curating possible new glitch classes. 
The Gravity Spy Zooniverse platform includes a forum where volunteers can discuss and ask questions of experts.
In addition to the interviews, we extracted all comments from the Gravity Spy forum, and used these to supplement interviewees' stories, focusing on the collaboration, articulations of work and how they coordinate action through the Gravity Spy's discussion interfaces. 

\subsubsection{Citizen-science volunteer investigations}

\begin{figure}
    \centering
    \includegraphics[width=0.75\textwidth]{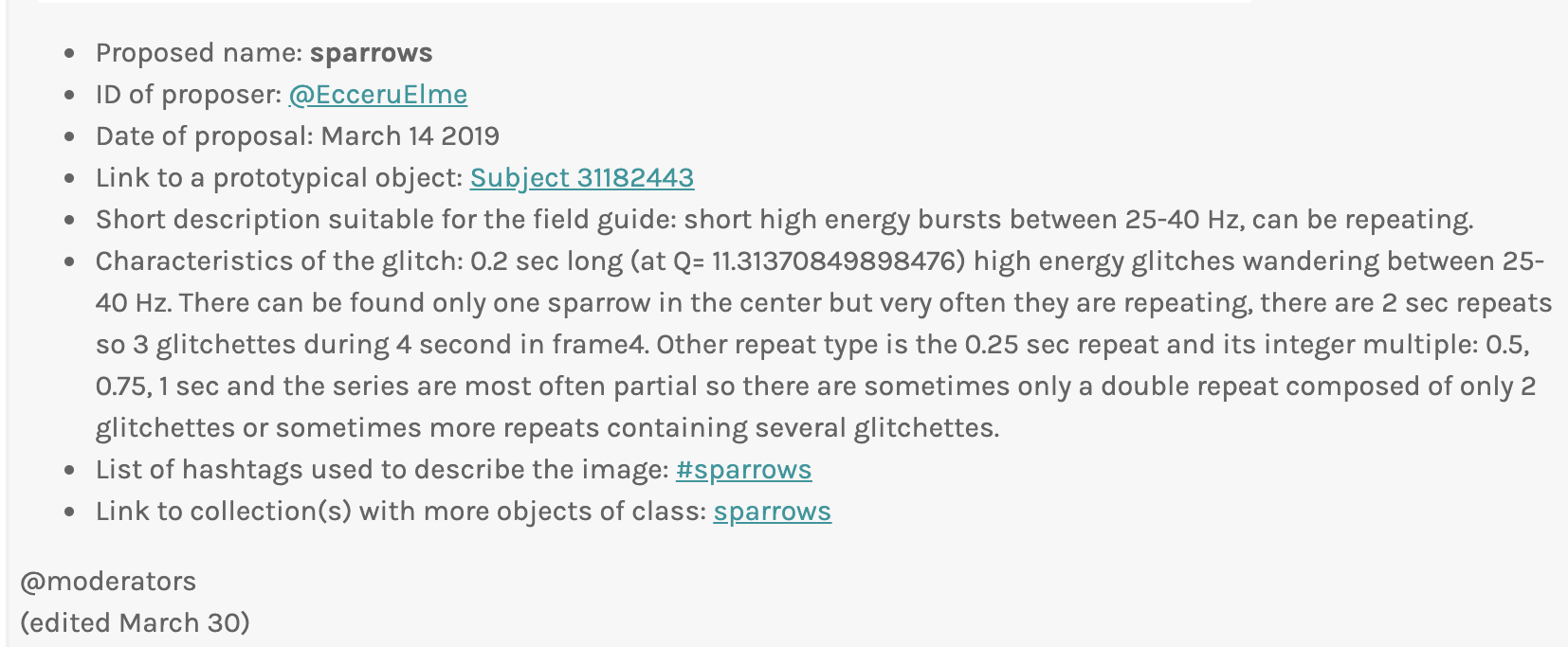}
    \caption{An example of a glitch proposal for a Sparrow class, which would eventually become the Crown class. 
    After collecting volunteer input, moderators submit proposals and include: a link to an exemplary image, a short description of the glitch class, a link to the hashtag and link to collections containing $\sim75$ examples of the glitch class.}
    \label{fig:proposal}
\end{figure}

To propose a new glitch class, volunteers must show that their proposed class is distinct from existing classes and abundant enough in the LIGO data to warrant attention. 
They do so by developing a new glitch proposal following a template outlined by our science team (see Figure~\ref{fig:proposal}). 
Glitch proposals describe the morphological characteristics of a glitch and link to collections of exemplary subjects. 
Proposals are vetted by moderators (science-team members or experienced volunteers) prior to submission to ensure they include all the necessary information and to verify that the proposed class is novel. 
Significant work goes into developing these proposals including searching for unusual glitches while classifying; curating subjects using hashtags (\#) in forum discussions; building collections; using the similarity search tool; describing glitch characteristics in discussion posts, and exploring the LIGO aLog.\footnote{aLogs are public records of issues at the LIGO detector sites: Hanford aLog \href{https://alog.ligo-wa.caltech.edu/aLOG/}{alog.ligo-wa.caltech.edu/aLOG/} and Livingston aLog \href{https://alog.ligo-wa.caltech.edu/aLOG/}{alog.ligo-wa.caltech.edu/aLOG/}.} 
Interviews and trace-data analysis suggest that volunteers rely on a range of search strategies. 
For instance, one volunteer reported that they relied on their photographic memory and hashtags to find potential new glitches, while another volunteer combined hashtag searches, the similarity search and collections. 
This work often continues after the submissions of a proposal as volunteers wait for science-team feedback.

\begin{figure}
    \centering
    \includegraphics[width=0.75\textwidth]{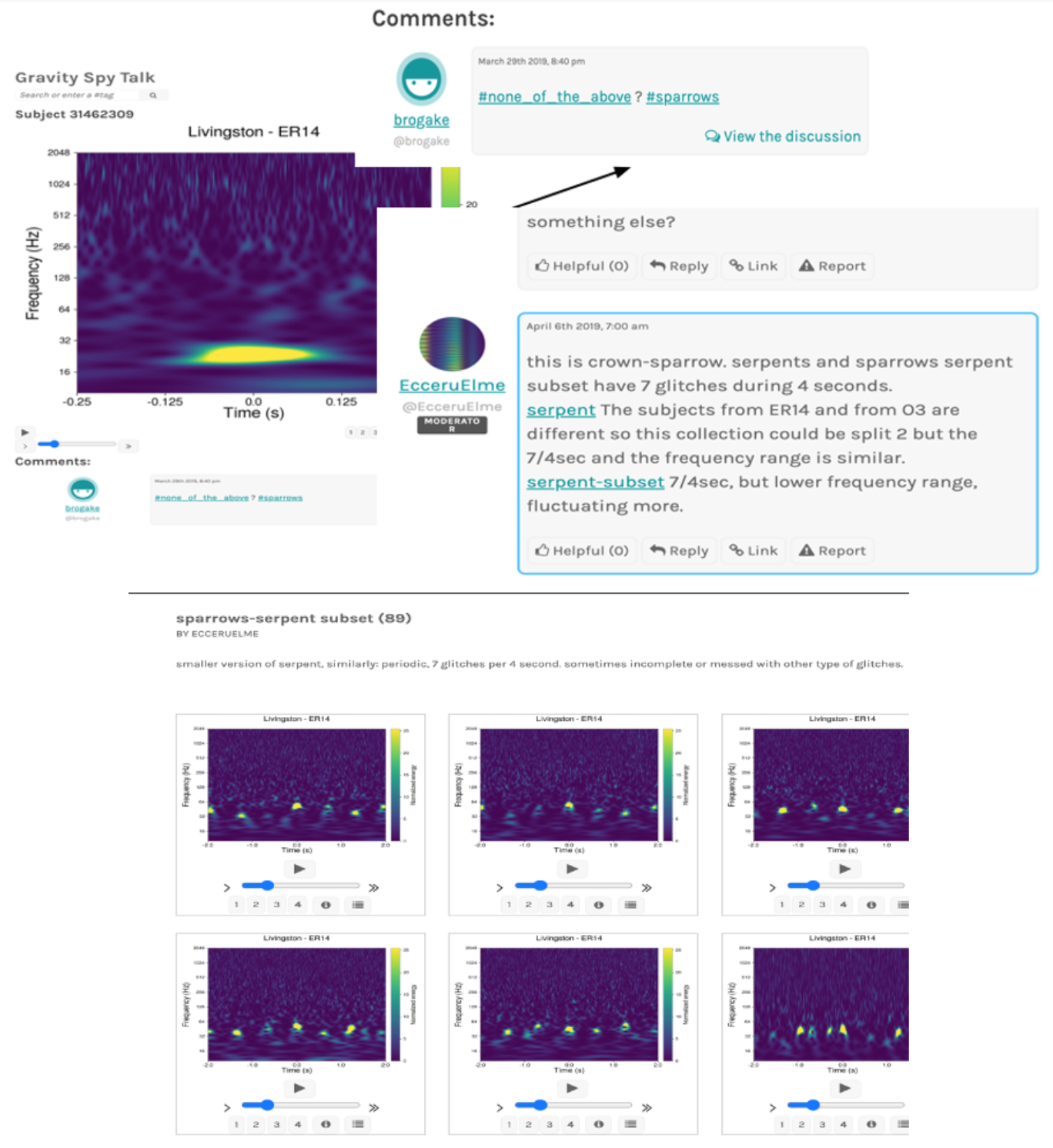}
    \caption{Instance of one volunteer curating glitches using hashtags and another articulating the differences between three glitch classes (\emph{top}) and a collection of subjects which the collection creator labelled sparrow-serpent subset (\emph{bottom}).
    There are $89$ subjects that the volunteer believes matches this class.
    ER14 and O3 refer to when the data were taken: Engineering Run 14 and Observing Run 3, respectively. 
    ER14 immediately preceded O3, and was a time to test the running of the detectors before making science-quality observations~\cite{Abbott:2020qfu}; the behavior of the detectors should be similar in ER14 and O3.}
    \label{fig:comments}
\end{figure}

Ideas for new glitch classes often emerge during volunteers’ regular classification activity where they come across (None of the Above) glitches that do not fit any of the known classes. 
However, there are tens of thousands of None of the Above subjects in the Gravity Spy data, and volunteers apply a range of strategies to group these subjects into coherent classes. 
Figure~\ref{fig:comments} illustrates the use of hashtags, comments and collections.

Many participants will mark an unknown subject with \#none\_of\_the\_above and sometimes propose a name for this potential new class (see top post in Figure~\ref{fig:comments}), which was part of the Sparrow proposal shown in Figure~\ref{fig:proposal}. 
Subsequently, other volunteers might comment on posts as we see in the second post in Figure~\ref{fig:comments}, where an experienced participant describes the differences between three glitches under investigation (crown-sparrow, sparrows and serpents) pointing to specific morphological features of each glitch e.g., ``sparrow serpent subset has 7 glitches during 4 seconds.'' 
Such comments may help newcomers learn which glitches are being curated and the characteristics of their shape~\cite{Jackson:2015}.

The Zooniverse platform offers the ability to create collections of subjects, which can be public or private; some volunteers will build collections of related glitch images which they populate with subjects they find during their regular classification, by searching for particular hashtags or keywords, adding glitches from other collections, or using the similarity search. 
Participants often use hashtags to search for subjects with similar names. 
Yet, we do not find common policies about tag use and how to reconcile conflicting naming practice. 
It is not uncommon to find deliberations about whether a subject belongs to e.g., a \#crown, \#sparrow or \#serpent class. 

Advanced volunteers rely on external resources such as the similarity search to curate subjects~\cite{Oesterlund:2014}. 
The similarity search tool can be used to display subjects which appear most similar to a prototypical glitch. 
This process supplements the manual and serendipitous search for similar subjects. 
The output of the similarity search includes supplementary data e.g., peak frequency and duration, which volunteers use to determine whether a subject fits within a collection. 
As evidenced in the glitch proposals, volunteers download and analyze supplementary data to determine whether subjects returned from the search fit within the numeric bounds of subjects already within the collection. 
%Those that fall within those bounds would then be included in the volunteers’ collection.  
Additional outputs of the similarity search are also used to provide evidence of the glitches and temporal prevalence. 
Distributional charts (another output of similarity search) were posted and synthesized by a volunteer who noted, 
``(1) crown cluster started around mid-March 2019, (2) \ldots{} the same problem was present already in mid-December 2018, (3) invisible cluster (1e+20 subjects, not on Gravity Spy) might be similar, and (4) between mid-February 2019 and beginning of July 2019 there seem to be two larger time clusters of crowns (couldn't feel the gap yet)''.\footnote{Glitches uploaded to Zooniverse have a numerical identification; for glitches in the Gravity Spy database yet to be uploaded to Zooniverse, the similarity search returns an identification number 1e+20.} 
Here the volunteer notes that Crown glitches become more common after the start of O3, that the machine-learning algorithm found them similar to pre-existing Scattered Light glitches, and that the prevalence of Crowns changes with time (following the prevalence of ground motion at the sites).
 
Finally, some volunteers reference aLogs in comments. 
Written by LIGO scientists, aLogs provide insights into the source of glitches. 
One volunteer composed an archaeology of aLog posts related to a potential new glitch named Crown.
 \begin{quote}
    \textbf{Summary:} Shaker injections at End-Y produce noise similar to `daytime noise' (noise that correlates with increased anthropogenic motion during daytime).

    \textbf{Characteristic shape:} As noted by Josh and Andy in alog 44803, the daytime noise that we see when anthropogenic ground motion is increased has a characteristic shape with both `slow' and `fast' arches. There are a few examples in the linked alog, and the first attached image file shows this same characteristic shape during a time when a train was passing.
\end{quote} 
These posts help to inform volunteers on potential details to look for in assembling collections, and also give an indication of changes in the state of the detectors which could impact the glitches present in the data. 
This aLog~\cite{alog:44803} points out the source of the Crown glitch at the LIGO detector site appearing in the Gravity Spy data and its characteristic shape---this volunteer-curated class is equivalent to the Fast Scattering glitch class.

After compiling evidence in a glitch proposal, volunteer suggestions for new glitch classes are reviewed by the LIGO detector-characterization experts. 
The experts check for possible origins of the glitch, and provide feedback after evaluating the proposal. 
In some cases, the proposed class may be a subset of an existing class or may be extremely localized in time (for example, the proposed Falcon class only occurred in a two-hour period on 20 June 2017 when there was an unreplicated issue with the Hanford detector), and so a new class may not be justified. 
If a new prevalent glitch class is identified, the volunteer-curated collection becomes the seed of a training set for the CNN.

\subsubsection{Volunteer-selected training set}

The Volunteer O3 training data set contains $9496$ training examples distributed over $23$ classes, with $265$ Crown examples: 
to aid the comparison between the proposed Fast Scattering and Crown classes, we include the expert-curated Low-frequency Blip examples, the additional Tomte examples and remove the None of the Above class for this model too. 
As for the Expert training set, the machine-learning algorithm was trained on this data set for a total of $100$ epochs, with a $87.5$:$12.5$ split between training and validation set. 
At the end of the training, the training and validation set accuracies were $99.2$\% and $97.5$\% respectively.

\section{Results}

Following the methods discussed above,
we have updated the Gravity Spy machine-learning algorithm using two new training sets. 
The first training set was assembled by LIGO detector characterization experts (Sec.~\ref{sec:expert}) using knowledge of the status of the detectors; this included two new classes: Fast Scattering and Low-frequency Blip. 
The second training set was assembled by citizen-science volunteers (Sec.~\ref{sec:volunteer}) following their investigations of spectrograms; this included the new class of Crown. 
We identify the proposed Fast Scattering and Crown classes as representing the same noise source caused by ground-motion at the detector site.
Using the two versions of the retrained machine-learning algorithm, we analyzed the O3 data from Livingston and Hanford to classify the glitch populations, and compared the models derived from the Expert and Volunteer training sets.

\subsection{Performance of the O3 models}

\begin{figure}
    \centering
    \includegraphics[width=0.9\textwidth]{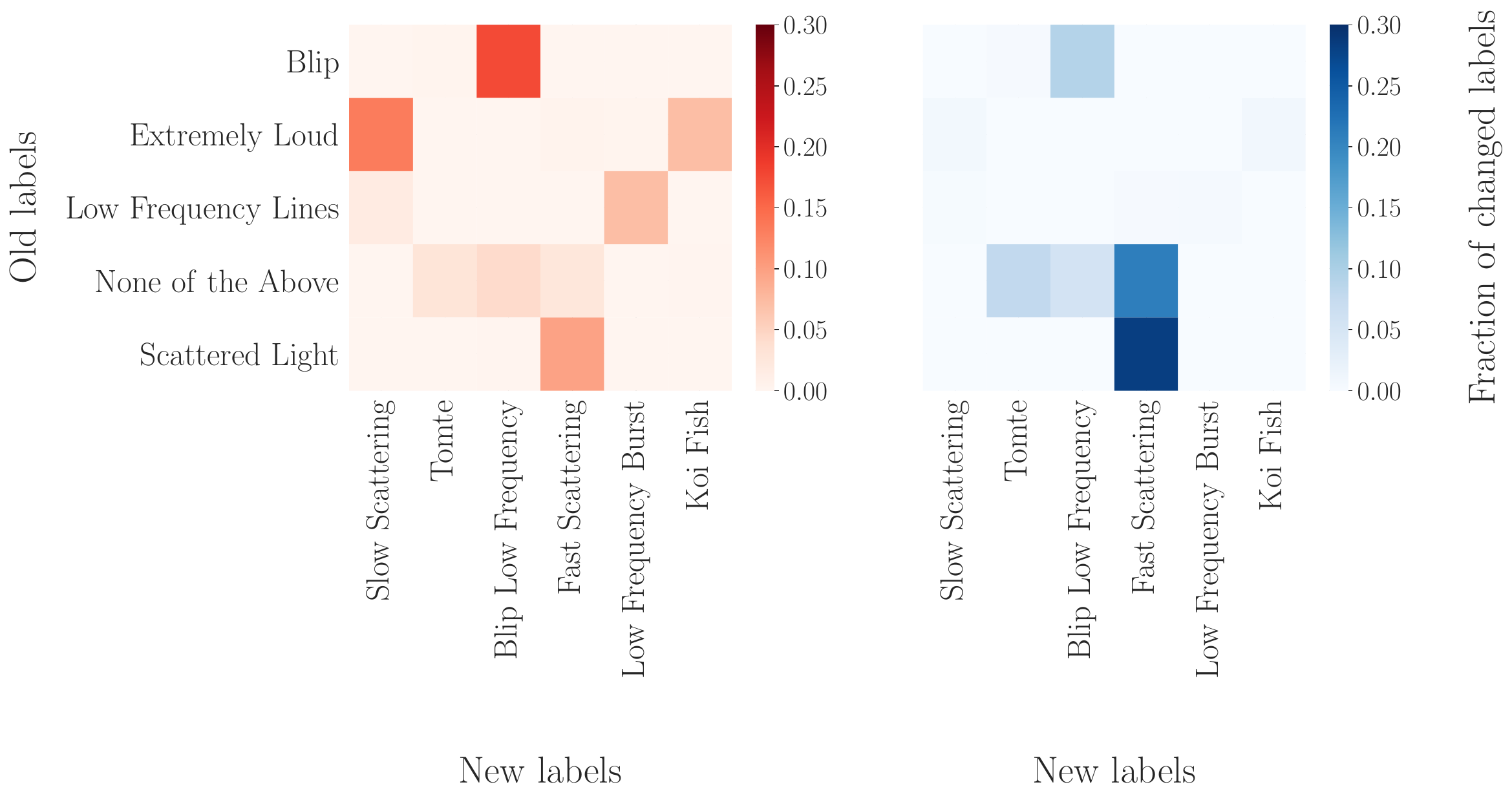}
    \caption{  Heatmaps showing changes of labels following reclassification by the Expert model for Hanford (\emph{left}) and Livingston (\emph{right}) data. 
    The change of labels was restricted to a handful of classes, and we only show the classes that contributed to at least $5 \%$ of the changes in labels on reclassification.}
    \label{fig:heatmap}
\end{figure}

Before reclassifying the O3 data with the new models, we need to check that the newly trained model successfully identifies the new classes (Fast Scattering and Low-frequency Blips for the Expert model) with high confidence levels. 
To ensure this, we reclassified $100$ triggers labeled as Scattered Light by the previous model on days with high ground motion in the anthropogenic band and with $Q$-value between $8$ and $14$. 
The selected range of $Q$-values ensured that they were Fast Scattering, and the new Expert model classified all of them as such. 
Similarly, we reclassified $80$ Blips with peak frequency between $10~\mathrm{Hz}$ and $50~\mathrm{Hz}$, and the new model classified $79$ of them as Low-frequency Blips and one as Tomte. 
We visually inspected all the reclassified triggers as a further measure of confidence. 
Correctly recognizing the newly added glitch categories is not enough: the new model should also not misclassify other glitch categories.

To probe more in-depth and to check the overall performance of this model, we reclassified $20\%$ of O3 data at Hanford and Livingston with the new model to determine:
\begin{enumerate}
    \item Is there a large change in the confidence assigned to data by the new model compared to the old model?
    \item What percentage of data is labelled with a new glitch category?
    \item What is the overall movement of labels from the old classification to the new classification?
\end{enumerate}
A comparison of the confidence distribution of the reclassified sample with the old classification did not show any significant difference for either of the detectors. 
At Livingston, the new model classified $42\%$ of the triggers in the sample with a new label, while at Hanford, only $11\%$ of the data was given a new glitch category.
This is expected since Fast Scattering is more frequent at Livingston, and for reasons we do not yet understand, we have observed a higher rate of Low-frequency Blips there. 

Figure~\ref{fig:heatmap} illustrates the main changes in glitch classification between the old and new Expert models. 
As expected, the majority of newly classified Fast Scattering at both detectors was labelled earlier as either Slow Scattering (previously named Scattered Light) or None of the Above. 
Similarly most of the Low-frequency Blips were previously classified as either Blips or None of the Above. 
At Hanford, some of the triggers previously classified as Extremely Loud are now labelled as Slow Scattering by the new model. 
Extremely Loud does not have a typical morphology similar to Slow Scattering, being more defined by its high SNR, but visual inspection of the reclassified examples confirm them as Slow Scattering. 
As shown in Figure~\ref{fig:glitches_combined}, glitches with a lot of saturation in the time--frequency plane between $10$--$1000~\mathrm{Hz}$ are classified as Extremely Loud. 
A number of Koi Fish glitches occurring within a short duration from each other can saturate the time--frequency plane as well, and hence impersonate Extremely Loud glitches. 
Due to this thin barrier between the two glitch categories from a morphological point of view, the movement from Extremely Loud to Koi Fish at Hanford is not surprising.
The results indicate that the updated model is classifying glitches as expected.

A summary of the most prevalent glitches, as classified by the old and new models is presented in Table~\ref{tab:top}. 

\begin{table*}
\caption{Most common glitch classes in LIGO Livingston (\emph{top}) and LIGO Hanford (\emph{bottom}) O3 data as labelled by the old and new (Expert and Volunteer) Gravity Spy models with confidence above $90\%$. 
%, and the total number of glitches classified with above $90\%$ confidence out of $289058$ for Livingston and $147818$ for Hanford. 
Percentages indicate the fraction of all glitches analyzed from the given observatory. 
The Scattered Light class was renamed Slow Scattering in the new models. 
The Expert model introduces the new classes Fast Scattering while the Volunteer model introduces Crown. 
Both of the models include the new Low-frequency Blip class, and neither includes a None of the Above class.
}
\label{tab:top}
\centering
%\begin{tabular}{c c c c@{\extracolsep{8pt}} c c c}
%\hline
% & \multicolumn{3}{c}{Livingston} & \multicolumn{3}{c}{Hanford} \\
%\cline{2-4} \cline{5-7}
%Rank & Old & Expert & Volunteer & Old & Expert & Volunteer \\
%\hline\hline
%\multirow{2}{*}{1} & Scattered Light   & Fast Scattering    & Slow Scattering & Scattered Light     & Slow Scattering & Slow Scattering  \\
%                   &  $40.32\%$        &  $27.13\%$         & $31.54\%$ & $46.60\%$           &  $46.95\%$          & $52.87\%$  \\
%\multirow{2}{*}{2} & Tomte             & Slow Scattering    & Crown & Low-frequency Burst & Low-frequency Burst & Low-frequency Burst  \\
%                   &  $13.91\%$        &  $23.22\%$         &$23.97\%$  &  $13.29\%$          &  $15.95\%$          & $14.55\%$ \\
%\multirow{2}{*}{3} & None of the Above & Tomte              & Tomte & Extremely Loud      & Extremely Loud & Extremely Loud  \\
%                   &  $12.17\%$        &  $19.31\%$         & $18.16\%$ &  $10.96\%$          &  $8.99\%$           & $9.00\%$  \\
%\multirow{2}{*}{4} & Blip              & Low-frequency Blip & Low-frequency Blip & Koi Fish            & Koi Fish  & Koi Fish  \\
%                   &  $5.88\%$         &  $7.67\%$          & $6.77\%$ &  $7.24\%$           &  $6.94\%$           & $6.76\%$  \\
%\multirow{2}{*}{5} & No Glitch         & Extremely Loud     & Extremely Loud & Blip & Blip & Blip  \\
%                   &  $5.18\%$         &  $3.58\%$          & $3.57\%$ &  $6.86\%$           &  $4.94\%$           & $4.21\%$  \\
%\hline
%\end{tabular}
\begin{tabular}{c c c c}
\hline
\multicolumn{4}{c}{Livingston} \\
%\hline
\cline{2-4}
Rank & Old & Expert & Volunteer \\
\hline\hline
\multirow{2}{*}{1} & Scattered Light   & Fast Scattering    & Slow Scattering    \\
                   &  $40.3\%$        &  $27.1\%$         &  $31.5\%$         \\
\multirow{2}{*}{2} & Tomte             & Slow Scattering    & Crown              \\
                   &  $13.9\%$        &  $23.2\%$         &  $24.0\%$         \\
\multirow{2}{*}{3} & None of the Above & Tomte              & Tomte              \\
                   &  $12.2\%$        &  $19.3\%$         &  $18.2\%$         \\
\multirow{2}{*}{4} & Blip              & Low-frequency Blip & Low-frequency Blip \\
                   &  $5.9\%$         &  $7.7\%$          &  $6.8\%$          \\
\multirow{2}{*}{5} & No Glitch         & Extremely Loud     & Extremely Loud     \\
                   &  $5.2\%$         &  $3.6\%$          &  $3.6\%$          \\
\hline
Total             & $195445$           & $203469$           & $216423$           \\
\hline\hline
\multicolumn{4}{c}{Hanford} \\
%\hline
\cline{2-4}
Rank & Old & Expert & Volunteer \\
\hline\hline
\multirow{2}{*}{1} & Scattered Light     & Slow Scattering     & Slow Scattering     \\
                   &  $46.6\%$          &  $46.9\%$          &  $52.9\%$          \\
\multirow{2}{*}{2} & Low-frequency Burst & Low-frequency Burst & Low-frequency Burst \\
                   &  $13.3\%$          &  $15.9\%$          &  $14.6\%$          \\
\multirow{2}{*}{3} & Extremely Loud      & Extremely Loud      & Extremely Loud      \\
                   &  $11.0\%$          &  $9.0\%$           &  $9.0\%$           \\
\multirow{2}{*}{4} & Koi Fish            & Koi Fish            & Koi Fish            \\
                   &  $7.2\%$           &  $7.0\%$           &  $6.8\%$           \\
\multirow{2}{*}{5} & Blip                & Blip                & Blip                \\
                   &  $6.9\%$           &  $4.9\%$           &  $4.2\%$           \\
\hline
Total              & $125256$            & $121632$            & $121113$            \\
\hline
\end{tabular}
\end{table*}

\subsection{Fast Scattering at LIGO Livingston}

\begin{figure}
    \centering
    \includegraphics[width=0.85\textwidth]{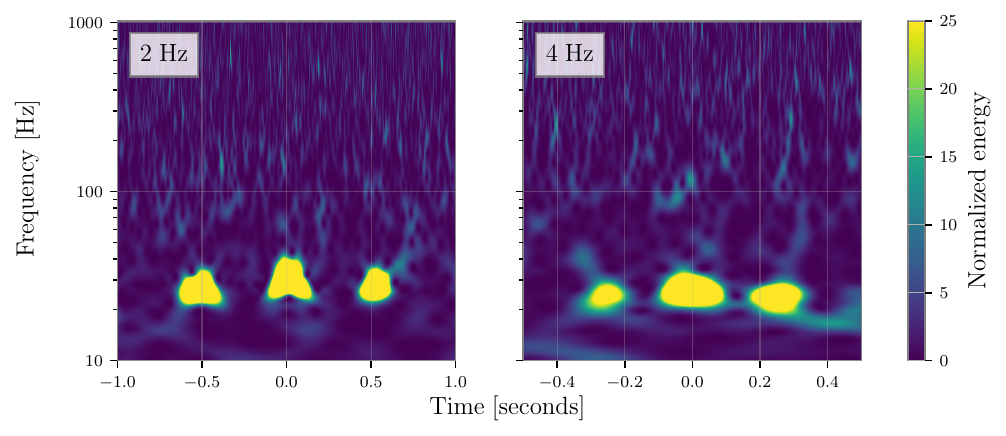}
    \caption{Spectrograms showing examples of the two subclasses of Fast Scattering.
    \emph{Left:} The $2~\mathrm{Hz}$ Fast Scattering is associated with an increase in microseismic ground motion ($0.1$--$0.3~\mathrm{Hz}$) at the observatory site. 
    \emph{Right:} The rate of $4~\mathrm{Hz}$ Fast Scattering goes up with an increase in the anthropogenic band ($1$--$6~\mathrm{Hz}$) caused due to trains, thunderstorms or human activity near the instrument. 
    Each subclass of Fast Scattering is made up of (a variable number) of individual bursts of noise which appear as blobs or arches in the spectrogram. }
    \label{fig:scan_fs}
\end{figure}

The primary motivation behind the retraining and reclassification was to generate a large data set for improved noise and statistical analysis, and this abundance of data provided us the resource to achieve this.  
We performed detailed investigations using the results of the Expert model; using this we found $55279$ glitches labelled as Fast Scattering with confidence above $90\%$ at Livingston. 
To study the origins of Fast Scattering, we focused on the month of February 2020 when $14536$ triggers were classified as Fast Scattering with $\geq 90\%$ confidence. 
Before this reclassification, we were aware of the $4~\mathrm{Hz}$ Fast Scattering that correlates strongly with ground motion in the anthropogenic $1$--$6~\mathrm{Hz}$ band. 
In the reclassified data set, we also found a strong presence of a $2~\mathrm{Hz}$ Fast Scattering population with a dependence on ground motion in $0.1$--$0.3~\mathrm{Hz}$ microseismic band. 
Fast Scattering glitches associated with these two distinct types of ground motion are shown in Figure~\ref{fig:scan_fs}. 
Ocean waves in the Gulf of Mexico are the primary source of microseismic noise at Livingston which is usually associated with Slow Scattering~\cite{Soni:2020rbu}.  
The microseismic coupling (and the reduction in Slow Scattering discussed next) explains the overall high rate of Fast Scattering in February, as microseismic ground motion is higher at Livingston during winter.

The LIGO detectors use a four-level suspension used to reduce the effects of ground vibration~\cite{Matichard:2015eva}. 
This consists of two parallel chains: the main chain, which includes the test mass as its bottom layer, and the reaction chain (RC), which provides an isolated set of masses that can apply control signals to the main chain~\cite{{Aston:2012ona,TheLIGOScientific:2014jea}}. 
During an increase in ground motion in the earthquake ($0.03$--$0.1~\mathrm{Hz}$) and microseismic ($0.1$--$0.3~\mathrm{Hz}$) bands, the relative motion between these two chains increases. 
This relative motion was observed to be highly correlated with an increase in the rate of Slow Scattering glitches.
To try to mitigate the impact of ground motion, a technique known as RC tracking was implemented at the detectors in the first half of January 2020~\cite{Soni:2020rbu}. 
RC tracking reduces the relative motion between the two chains by sending a part of the control drive originally applied to the penultimate stage, to the top (R0) stage of the suspension.
This reduced the rate of Slow Scattering by a factor of $\sim100$ for microseismic ground motion above $1000~\mathrm{nm\,s^{-1}}$ at Hanford and $1500~\mathrm{nm\,s^{-1}}$ at Livingston. 
The reduction in noise could have contributed to an increased visibility of $2~\mathrm{Hz}$ Fast Scattering, which may have been previously hidden underneath (louder) Slow Scattering.

\subsection{Comparing Expert and Volunteer}

In order to analyze the similarities and differences between the Expert model's Fast Scattering and the Volunteer model's Crown, we ask:
\begin{enumerate}
    \item What is the distribution of glitches classified as Fast Scattering and Crown glitches by our two models?
    \item Is the rate of Fast Scattering and Crown impacted by similar physical conditions near the detector?
\end{enumerate}
Due to an increase in microseismic as well as anthropogenic ground motion near Livingston, February 2020 registered a large number of glitches classified as Fast Scattering and Crown by our two models, and we used this sample to answer our questions.

\begin{figure}
\captionsetup[subfigure]{}
   \centering
    \begin{subfigure}[b]{0.45\textwidth}
        \centering
         \includegraphics[width= \textwidth]{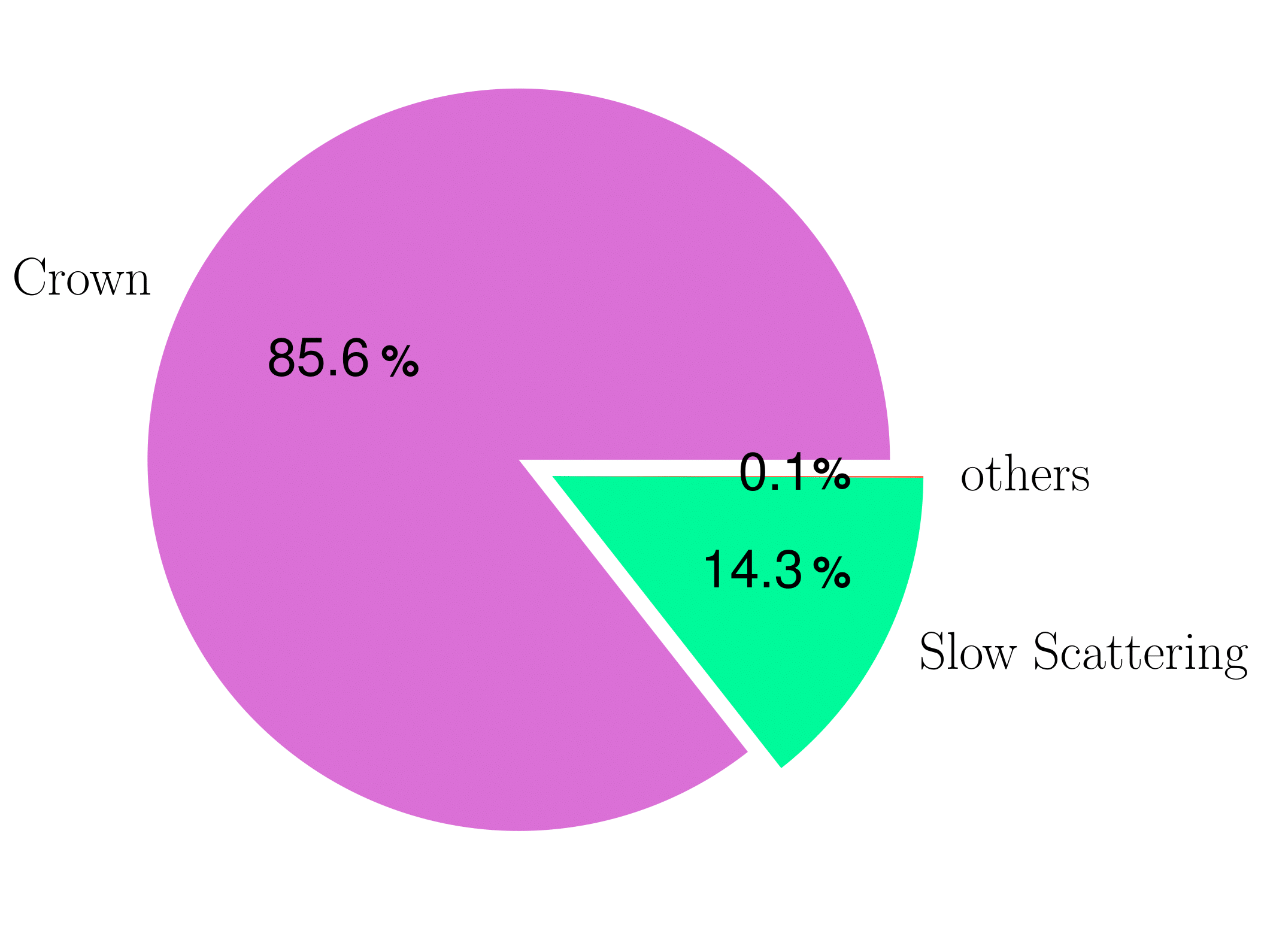}
         \caption{Volunteer model classification of Fast Scattering data.}
         \label{fig:vol_comp}
    \end{subfigure}
  \hfill
    \begin{subfigure}[b]{0.43\textwidth}
        \centering
         \includegraphics[width =\textwidth]{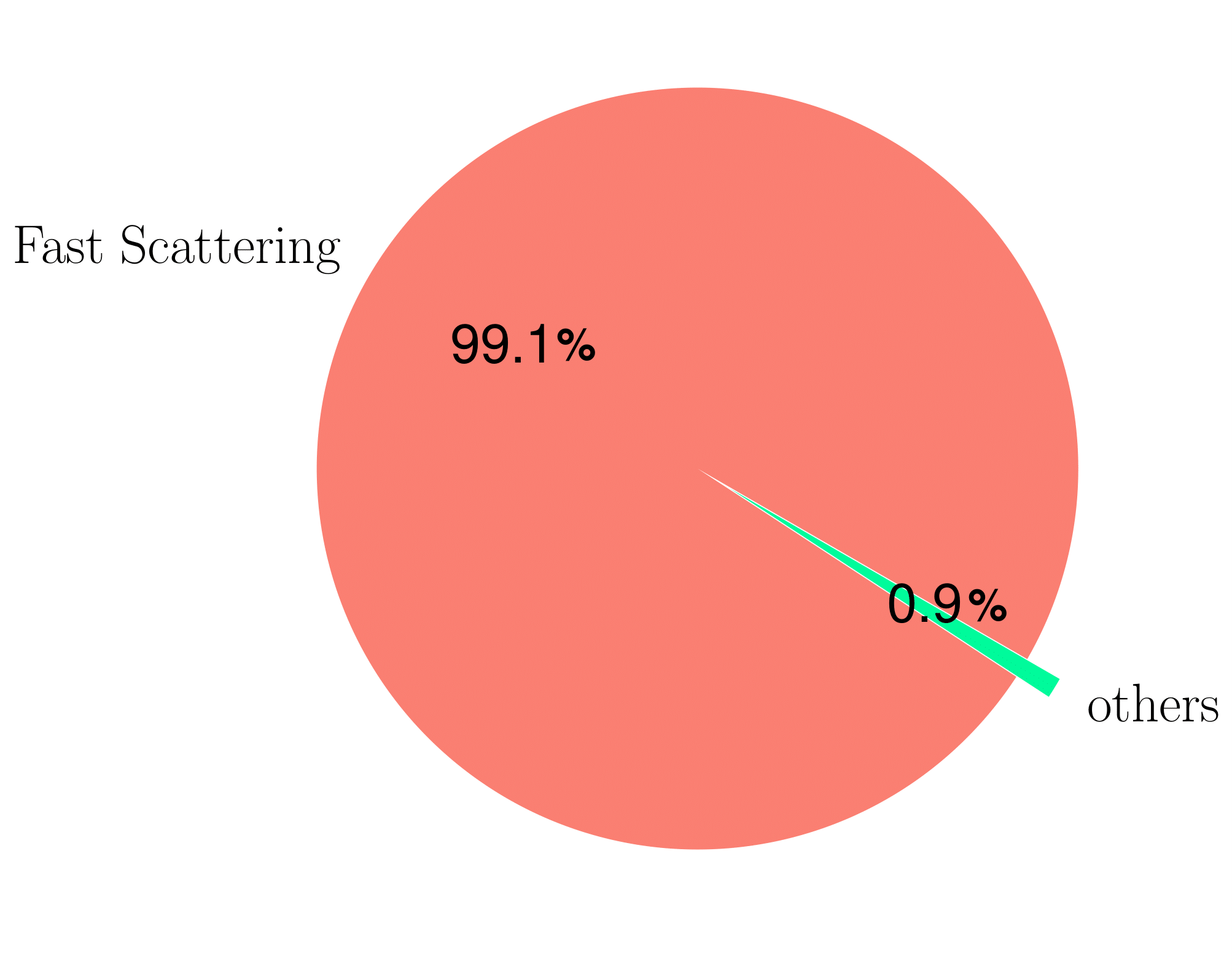}
         \caption{Expert model classification of Crown data.}
         \label{fig:exp_comp}
    \end{subfigure}
    \caption{Consistency of the Fast Scattering and Crown classes. \emph{Left}: Fast Scattering glitches classifications under the Volunteer model. 
    \emph{Right}: Crown glitches classifications under the Expert model. 
    The glitches have a minimum SNR threshold of $7.5$ and a minimum confidence of $0.9$. 
    }
    \label{fig:vol_comparison}
    
\end{figure}

To check the consistency of the Crown and Fast Scattering labels, we reclassified the glitches labelled as Fast Scattering with the Volunteer model and glitches labelled as Crown with the Expert model. 
This comparison is shown in Figure~\ref{fig:vol_comparison}. 
The Expert model classifies almost all ($>99\%$) the Crown glitches as Fast Scattering. 
The Volunteer model classifies the majority ($\sim85\%$) of Fast Scattering glitches as Crown, but also classifies a significant minority ($\sim14\%$) as Slow Scattering. 
We inspected the spectrograms of a sample of these glitches labelled as Slow Scattering and confirmed they belong to the Fast Scattering/Crown category. 
This confusion originates from the features common to Fast Scattering/Crown and Slow Scattering which can make it hard for the CNN to distinguish them. 
It was this similarity which led to many of the Fast Scattering/Crown glitches being classified as Slow Scattering using the old model. 
Overall, both models do primarily identify the same types of glitch.

\begin{figure}[h]
\captionsetup[subfigure]{}
   \centering
    \begin{subfigure}[b]{0.4\textwidth}
        \centering
         \includegraphics[width= \textwidth]{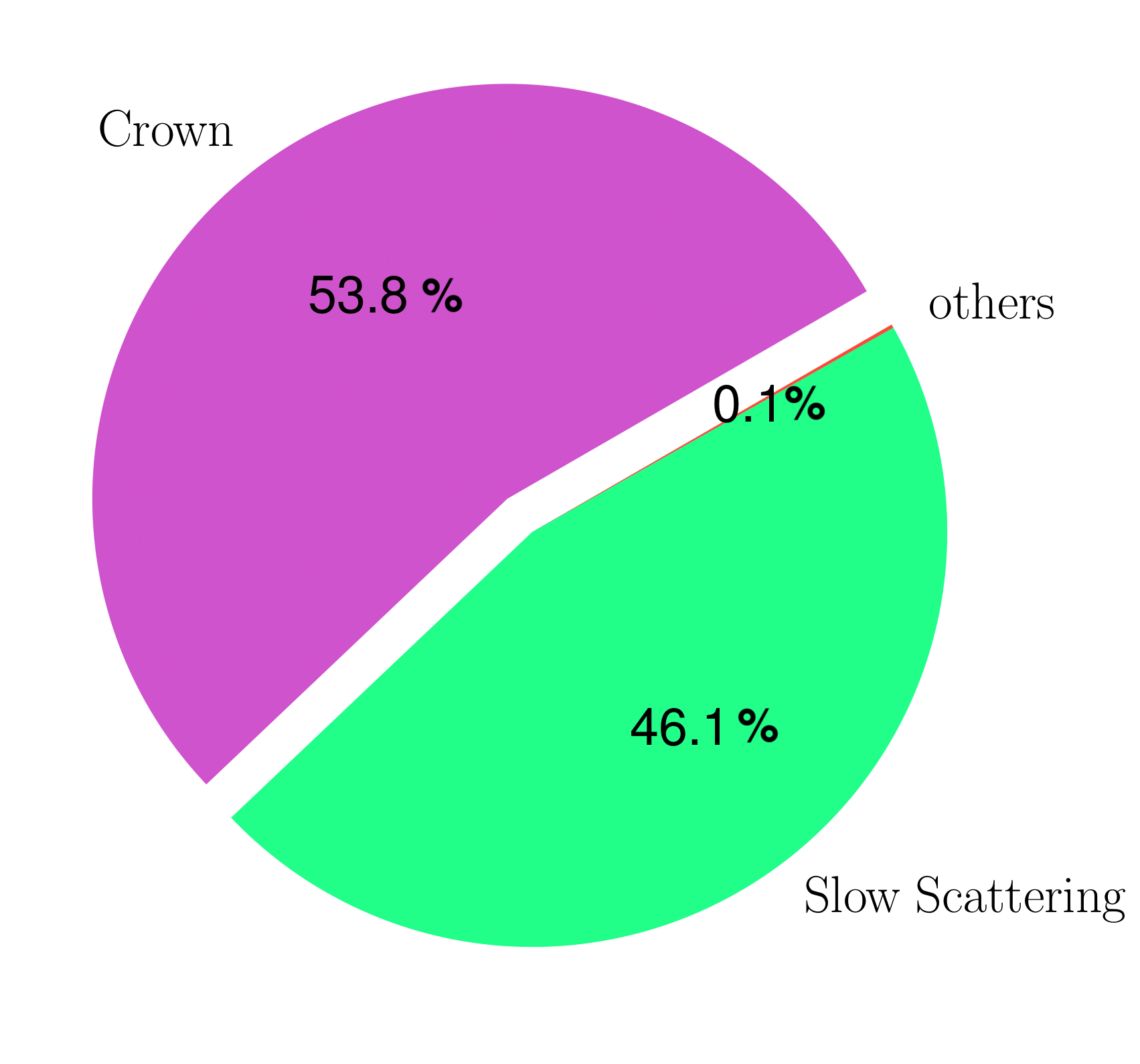}
         \caption{Volunteer model classification of Fast Scattering with SNR $> 20$.}
         \label{fig:vol_comp20}
    \end{subfigure}
  \hfill
    \begin{subfigure}[b]{0.43\textwidth}
        \centering
         \includegraphics[width =\textwidth]{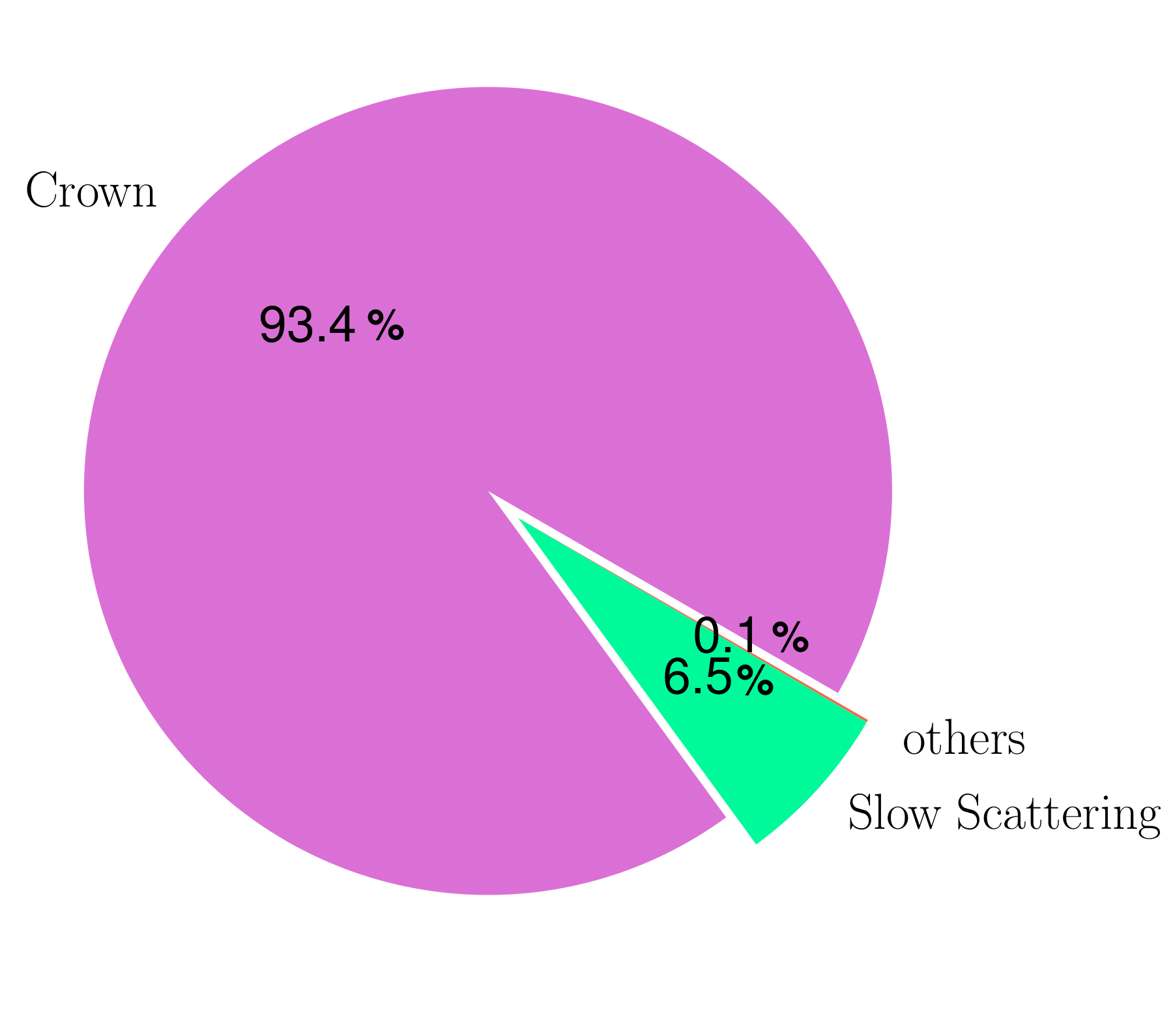}
         \caption{Volunteer model classification of 
         \newline Fast scattering with $7.5<$ SNR $< 10$.}
         \label{fig:vol_comp8}
    \end{subfigure}
    \caption{Consistency of the Fast Scattering and Crown classes for different SNR thresholds: Fast Scattering classifications using the Volunteer model. 
    The fraction of Fast Scattering glitches labelled as Slow Scattering instead of Crown increases with SNR.
    }
    \label{fig:vol_compSNR}
    
\end{figure}
The differences between the Expert and Volunteer models can be explained due to the differences in the training sets. 
%Figure~\ref{fig:ground_motion} highlights that while both Fast Scattering and Crown are correlated with ground motion, there are fewer high-SNR Crown glitches. 
Figure~\ref{fig:vol_compSNR} shows the classification of Fast Scattering glitches by the Volunteer model with different SNR thresholds. 
The fraction labelled as Slow Scattering increases with SNR. 
%\subsubsection{Expert and volunteer training sets} 
The Volunteer model classifies few high-SNR glitches as Crown. 
While the Expert training set contains $400$ Fast Scattering examples randomly selected from the O3 data between 1 February 2020 and 1 March 2020, the majority of the Volunteer training set ($221/265$) comes from the Engineering Run (ER14) during March 2019, just before the beginning of O3 in April 2019. 
Glitches from ER14 were uploaded to Zooniverse before those from O3, and so the first examples of Fast Scattering/Crown seen by volunteers were from this period.
We do not expect the behavior or the morphology of noise to change significantly from ER14 to O3, but we do find that there is a difference in the SNR of Fast Scattering/Crown glitches, mirroring the two training sets. 
The variation in Fast Scattering/Crown properties between ER14 and O3, while small, shows the importance of verifying if changes in the detectors have affected the form of glitches in the data.

\begin{figure}
    \centering
    \includegraphics[width=0.75\textwidth]{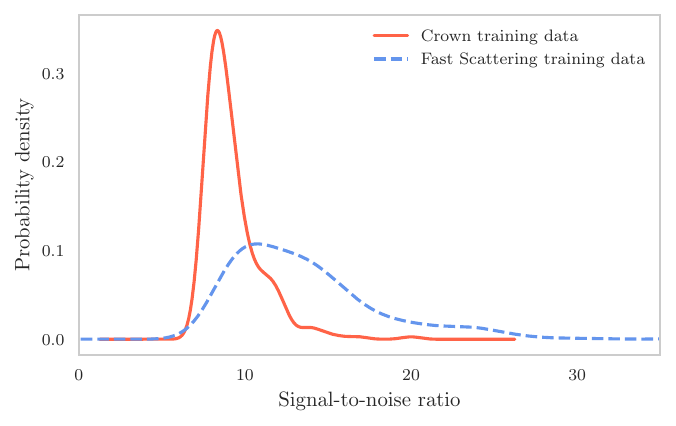}
    \caption{
    SNR distribution of Fast Scattering and Crown glitches in the Expert and Volunteer training sets, respectively. 
    The Crown training set has a narrower SNR distribution and peaks at a smaller value in comparison to the Fast Scattering training set.}
    \label{fig:density_volexp}
\end{figure}

Figure~\ref{fig:density_volexp} shows the SNR probability density of the Fast Scattering and Crown training data. 
The Fast Scattering training set covers a broader range of SNR values, with a symmetric $90\%$ range of $8.6$--$23.1$.  
In comparison, the Crown training data has a narrower SNR distribution with a $90\%$ range of $7.6$--$12.2$.
% This difference in the range of SNR of the individual training sets explains the disparity between the Expert and Volunters model evident in Figure~\ref{fig:vol_comparison} and Figure~\ref{fig:vol_compSNR}.
% The SNR distribution of the Volunteer training set has a mean of $9.24$ and a variance of $2.92$, whereas the respective values for the Expert training set are $13.57$ and $19.27$. 
%Also, the misclassification rate would have been similar across all SNR values if the glitch morphology was very different for the two training sets. 
The Expert training set covers a larger range of SNR values, and this difference explains the disparity between the Expert and Volunteer model evident in Figure~\ref{fig:vol_comparison} and Figure~\ref{fig:vol_compSNR}.   

The LIGO sites have a network of sensors to record their surrounding environments~\cite{TheLIGOScientific:2016zmo,Nguyen:2021ybi}. 
Ground motion (along the $X$, $Y$ and $Z$ directions) is measured by reference seismometers located at the two end stations and at the corner station of the LIGO interferometer~\cite{TheLIGOScientific:2014jea,Nguyen:2021ybi}. 
The output from these seismometers is then bandpassed into multiple frequency bands between $0.03~\mathrm{Hz}$ and $30~\mathrm{Hz}$. 
%The seismic noise plots in the multiple frequency bands are available on the LIGO Summary Pages, which are a set of web pages showing the state of interferometer for each day \cite{ligo_summary}.

\begin{figure}
    \centering
    \includegraphics[width=0.9\textwidth]{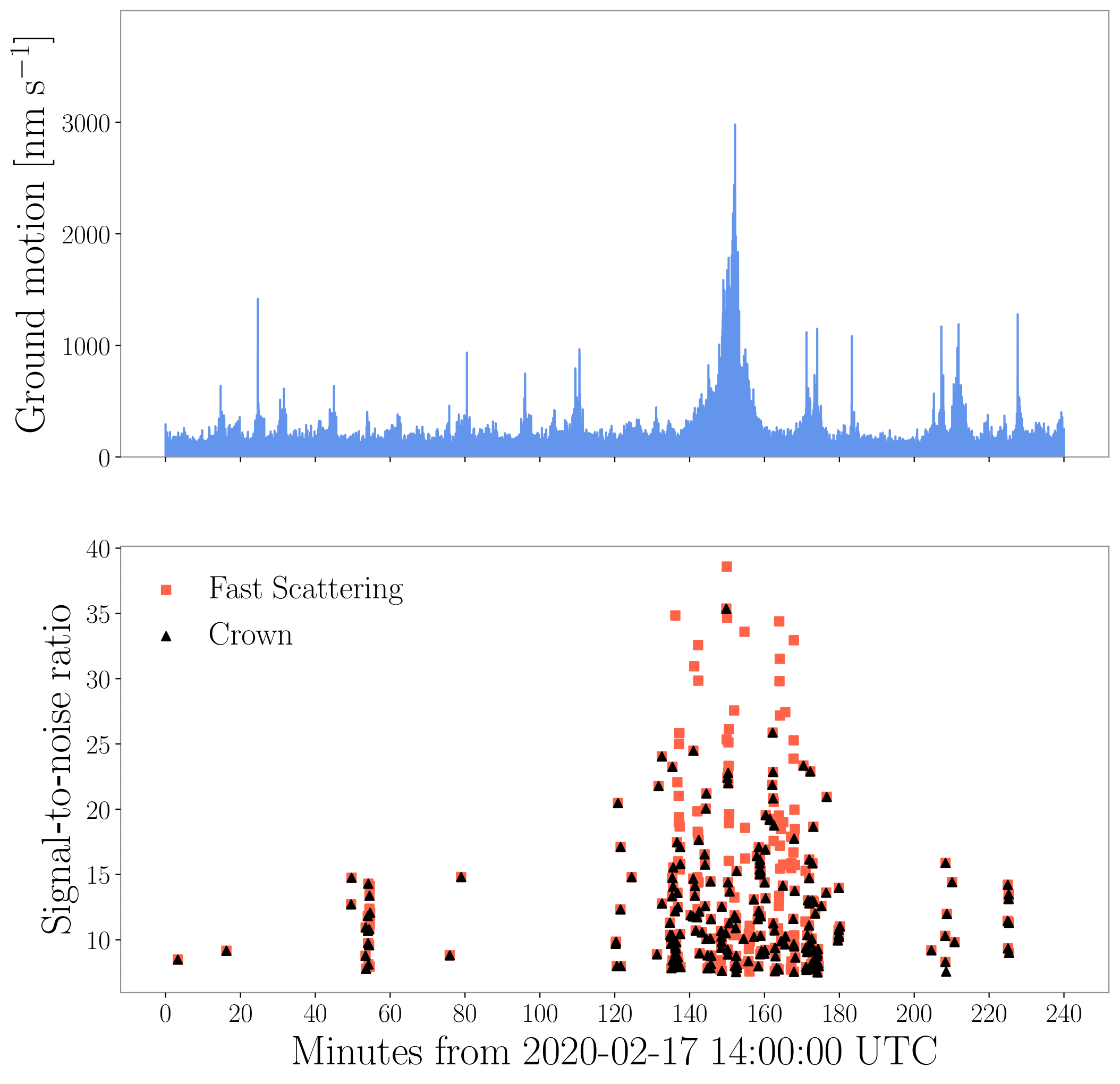}
    \caption{\emph{Top:} Ground motion recorded at the $Y$ end station of the LIGO Livingston Observatory, bandpassed between $3~\mathrm{Hz}$ and $5~\mathrm{Hz}$. 
    Trains on a track near this station regularly increase the ground motion in this anthropogenic band.
    \emph{Bottom:} SNR versus time plot of Fast scattering and Crown. 
    This shows an increase in the rate of both Fast Scattering and Crown glitches with an increase in the ground motion.
    %However, above SNR $\sim 25$, there are very few Crown glitches compared to Fast scattering.
    }
    \label{fig:ground_motion}
\end{figure}

To verify that Fast Scattering and Crown glitches have similar physical origins, we examine their rate as antropogenic ground motion varies. 
Several times a day, trains on a railway track near the $Y$ end of the Livingston detector cause anthropogenic-band ground motion. 
Figure~\ref{fig:ground_motion} shows a period of increased ground motion in the antropogenic band ($3$--$5~\mathrm{Hz}$) caused by a train on 17 February 2020 coincided with an increase in the rate of Fast Scattering and Crown glitches. 
The only significant difference between the two classes is observed at high SNR, as a consequence of the lack of high-SNR Crown glitches in the Volunteer training set. 
The correlation with ground motion shows that the noise recognized as Crown and Fast Scattering are affected by similar environmental factors. 
Correlating specific behaviors of transient noise such as its rate to environmental conditions is one of the most crucial steps in noise characterization.

\section{Discussion \& conclusions}

We have demonstrated how Gravity Spy can be used by both detector-characterization experts and citizen-science volunteers to efficiently identify new classes of glitch in the wealth of data provided by the LIGO gravitational-wave detectors. 
Examining the O3 data set, we have found two new families of glitches: Fast Scattering/Crown and Low-frequency Blip. 
% Using the Gravity Spy machine-learning algorithm trained with the detector-characterization curated training set classifies $55211$ and $1286$ glitches as Fast Scattering in O3 data with $90\%$ confidence from LIGO Livingston and LIGO Hanford, respectively, while $15614$ and $2467$ glitches are similarly classified as Low-frequency Blips. 
% Equivalently, using the citizen-science volunteer training set, the Gravity Spy machine learning algorithm classifies $51887$ and $581$ glitches as Crowns, plus $14670$ and $2635$ as Low-frequency Blips. 
%The new glitches are especially prevalent in O3 LIGO Livingston data.

We observe a significant difference between the glitch populations in LIGO Livingston and LIGO Hanford data. 
Using our Expert/Volunteer Gravity Spy machine-learning algorithm, $55211$/$51887$ glitches are classified as Fast Scattering/Crown with $90\%$ confidence in O3 LIGO Livingston data, while only $1286$/$581$ are classified as such in LIGO Hanford data. 
This highlights that while both observatories share the same base design~\cite{TheLIGOScientific:2014jea}, they differ in exact configuration and in their physical environments. 
Therefore, in future studies, it may be beneficial to tailor separate machine-learning algorithms to classify glitches in each detector. 
Since we expect that glitch classification will still require recognising similar features in time--frequency spectrograms, these observatory-specific algorithms could be built upon the existing Gravity Spy algorithm as an example of transfer learning~\cite{Coughlin:2019ref,Shao:2015,Tan:2018,Zhuang:2021}. 
Isolating new glitch classes enables progress in improving gravitational-wave detector data quality, and studying the glitch populations of different observatories individually may enable us to more precisely trace the origins of glitches.

We have shown that Fast Scattering/Crown glitches are related to ground motion at the detector sites. 
Through examining glitches classified as Fast Scattering, we have discovered $4~\mathrm{Hz}$ and $2~\mathrm{Hz}$ subclasses which have different dependencies on ground motion, which is a significant step towards improved noise characterization. 
Physical environment monitoring tests at the end and corner stations at both detector sites revealed high quality-factor resonances at frequencies close to $4~\mathrm{Hz}$ believed to be responsible for the $4~\mathrm{Hz}$ Fast Scattering observed during O3~\cite{alog:EX_reson_1,alog:EY_reson_1,alog:EY_reson_2,alog:corner_reson_1}. 
Scientists at both Livingston and Hanford have planned to damp this motion and re-evaluate the impact of ground motion on the detectors to see if data quality can be improved. 
As glitches caused by light scattering were a significant data-quality issue in O3, impacting the analyses of multiple detections~\cite{Abbott:2020niy}, mitigation of their causes is a priority ahead of the up-coming fourth observing run (O4)~\cite{Davis:2021ecd}.

The global detector network will continue to be enhanced to increase sensitivity to gravitational waves.
The A+ upgrades of the LIGO detectors include higher laser power, new test mass mirrors with lower thermal noise, balanced homodyne readout and frequency-dependent squeezing~\cite{Miller:2014kma}. 
Some of these upgrades will be completed before O4, increasing the binary neutron star inspiral ranges (an average distance a standard source can be detected~\cite{Chen:2017wpg}) from $\sim110$--$130~\mathrm{Mpc}$~\cite{Abbott:2020niy,Buikema:2020dlj} to $\sim160$--$190~\mathrm{Mpc}$~\cite{Abbott:2020qfu}. 
Other improvements to the detectors are planned to address data-quality issues: new baffles will be installed to reduce stray light~\cite{Buikema:2020dlj}, which should decrease the amount of scattered light, and the adoption of new seismic-isolation control systems should  improve the noise coupling from ground motion~\cite{Soni:2020rbu,Schwartz:2020pso}.
However, the process of detector calibration, upgrades and installing new hardware contribute to a change in the state of the system and may introduce new sources of transient noise. 
A gravitational-wave detector with enhanced sensitivity, and thus a lowered noise floor, may also witness new sources of noise that were not a cause of concern earlier. 
We may also observe the same transient noise but in a different frequency band~\cite{alog:3.3_Hz_fast_scatter}. 
Therefore, we expect that the identification of new glitch classes and their causes will remain a necessary task.

The similarity of the Fast Scattering and Crown classes identified independently by detector-characterization experts and citizen-science volunteers has provided a unique opportunity to study how the two groups work. 
We have shown that the crowd-sourcing potential of citizen-science projects does not have to be limited to simple classification tasks. 
Non-experts can explore complicated data sets and make impactful discoveries. 
We have found that volunteers engage in in-depth investigations, similar to professional scientists. 
They use a variety of tools in combination, and are capable of producing sophisticated analyses of their own design. 
While the Fast Scattering and Crown results are not identical, this is not surprising given that the volunteers lacked access to additional auxiliary information on the state of the detectors, such as the presence of ground motion around the time of the glitches, and that the volunteers created their training set using glitches from primarily ER14 rather than O3.
Our results suggest that providing volunteers with more data, as well as the tools to explore it, may lead to even greater scope for discovery.

The benefit of making data accessible to non-experts is that it can then be examined by many more people. 
Gravity Spy volunteers, like other citizen-science volunteers~\cite{Raddick:2013bca,Jones:2018}, are largely self-motivated, and like to work individually; however, expert feedback is appreciated and beneficial to their work. 
Therefore, we have found that it essential to provide (i) guidance to volunteers so that they can train themselves in the domain at their own pace (as through the Gravity Spy levels); (ii) tools to explore the data (such as the Gravity Spy similarity search~\cite{Coughlin:2019ref}), and (iii) the ability to interact with experts (as provided through evaluation of glitch proposals). 
These recommendations overlap with best practices advised for citizen-science projects to produce reliable results~\cite{Freitag:2006}. 
Previous studies have shown that citizen-science volunteers can produce high accuracy results, performing classifications comparable to professional scientists~\cite{Lintott:2008ne,Aceves-Bueno:2017,McCarthy:2021}, and even detect novel features in data~\cite{Lintott:2009wq,Schwamb:2012xp,Debes:2019}; here, we have shown volunteers can go beyond this, and identify new classes analogously to experts. 
When provided with suitable resources, citizen-science volunteers can make significant scientific discoveries based upon complex data sets. 
As science progresses to produce more data than can be examined by small science teams, the potential of citizen-science volunteers may enable even greater scientific return.

\ack{}

We thank the citizen-science volunteers of Gravity Spy who have contributed to the classifications of LIGO data. 
We are grateful to Laura Nuttall for comments on this manuscript, Michael Zevin for useful discussions,  and to the anonymous referees for their suggestions. 
Gravity Spy is partly supported by the National Science Foundation (NSF) award INSPIRE 15-47880. 
This work is supported by the NSF under grant PHY-1912648. 
SS is supported by NSF grant PHY-1806656.
CPLB is supported by the CIERA Board of Visitors Research Professorship. 
OP is supported by NSF grant PHY-1559694.
The authors gratefully acknowledge the support of the United States NSF for the construction and operation of the LIGO Laboratory and Advanced LIGO as well as the Science and Technology Facilities Council of the United Kingdom, and the Max-Planck-Society for support of the construction of Advanced LIGO. 
Additional support for Advanced LIGO was provided by the Australian Research Council. 
LIGO was constructed by the California Institute 
of Technology and Massachusetts Institute of Technology with funding from the NSF, and operates under cooperative agreement PHY-1764464. 
Advanced LIGO was built under award PHY-0823459.
LIGO was constructed by the California Institute of Technology and Massachusetts Institute of Technology with funding from the National Science Foundation and operates under Cooperative Agreement PHY-1764464. 
This work used computing resources at CIERA funded
by NSF grant PHY-1726951, and the computational resources and staff contributions provided for the Quest high performance
computing facility at Northwestern University which is
jointly supported by the Office of the Provost, the Office
for Research, and Northwestern University Information
Technology.
The authors are grateful for computational resources provided by the LIGO Laboratory and supported by National Science Foundation Grants PHY-0757058 and PHY-0823459. 
This document has been assigned LIGO document number \href{https://dcc.ligo.org/LIGO-P2100019/public}{LIGO-P2100019}.

%\textbf{Author contributions:}
%SS led the detector-characterization analysis, the analysis and curation of the data, and the drafting of the manuscript. 
%CPLB contributed to the conceptualization of this paper, the interpretation of the data, and the drafting of the manuscript.
%SBC supervized the analysis, contributed to the conceptualization of the paper, and assisted with drafting the manuscript.
%OP contributed to coordination of citizen-science volunteers' and detector-characterization analyzes, and validation of new glitch class proposals.
%AKK contributed to the development of the machine learning algorithms described in this paper.
%All authors contributed to editing the manuscript.

\bibliographystyle{iopart-num}
\bibliography{gravityspy.bib}

\end{document}